# Single Crystal Growth of Cuprate Superconductor

# $(Lu_{0.8}Nd_{0.2})Ba_2Cu_4O_8$ by KOH Flux Method

Hiroshi Hara[1,2*], Shintaro Adachi[1], Ryo Matsumoto[1,2], Aichi Yamashita[1,2], Hiroyuki Takeya[1]

and Yoshihiko Takano[1,2]

[1]*MANA, National Institute for Materials Science (NIMS), 1-2-1 Sengen, Tsukuba, Ibaraki*

*305-0047, Japan*

[2]*Graduate School of Pure and Applied Sciences, University of Tsukuba, 1-1-1 Tennodai,*

*Tsukuba, Ibaraki 305-8577, Japan*

**Abstract**

Single crystals of Nd-substituted $LuBa_2Cu_4O_8$ were successfully grown by the KOH flux method. The single phase of $Lu_{1-x}Nd_xBa_2Cu_4O_8$ [(Lu,Nd)124] formed at $x = 0.2$. The compound crystallized as orthorhombic *Ammm* structure with lattice constants of $a = 3.835(3)$ Å, $b = 3.879(2)$ Å and $c = 27.195(6)$ Å. Single crystal structural analysis demonstrated that the Nd ion partially occupied the Lu site. The (Lu,Nd)124 exhibited the superconducting transition at ~75 K in the magnetic susceptibility and resistivity. The $T_c$ and the $c$-axis of the (Lu,Nd)124 corresponded to Tm124. These results indicate that RE124 equivalent to the one composed of single RE element is obtained by combination of two RE elements.

It is well known that the Y site in $YBa_2Cu_3O_{7-\delta}$ (Y123) can be completely replaced by some of rare-earth (RE) elements and almost all RE123s show superconductivity at ~90 K.[1,2] Similarly, the Y site in $YBa_2Cu_4O_8$ (Y124) can also be replaced by several RE elements but the superconducting transition temperature ($T_c$) of RE124 varies between 70 K and 80 K.[3-7] Up to now, a series of RE124 has been obtained for RE = Sm, Eu, Gd, Dy, Ho, Er, Tm and Yb, and the $T_c$ of RE124 increases from 70 K to 80 K dependent on the decrease of the $c$-axis in this order.[3-11] In that sense, Lu124, consisting of the smallest RE ion of Lu, is expected to show the highest $T_c$ in the RE124 system since it has the smallest $c$-axis length.

Single crystals of RE124 for RE = Sm, Eu, Gd, Dy, Ho, Er, Tm and Yb were grown under ambient pressure by the potassium hydroxide (KOH) flux method.[8-11] However, Lu124 has not been synthesized yet due to the large lattice mismatch of the perovskite units caused by the small size of $Lu^{3+}$ ion, similar to the situation of Lu123.[12] If the lattice mismatch is reduced by partial substitution of Lu with a larger RE, the Lu124 structure can be stabilized. Here we report the successful growth of Lu124 single crystals by the KOH flux method with partial substitution of Lu site. We selected Nd as a substituting ion owing to the larger ionic radius of $Nd^{3+}$ (eight coordinates; 1.109 Å) than the smaller one of $Lu^{3+}$ (eight coordinates; 0.977 Å). Further, since Nd123 forms prior to formation of Nd124, single crystals of Nd124 have been hardly grown by KOH flux method.[8,11] Therefore, it is expected to reduce the lattice mismatch efficiently to stabilize the Lu124 structure and there is no risk of the contamination of Nd124 in the obtained samples.

Single crystal growth of $Lu_{1-x}Nd_xBa_2Cu_4O_8$ [(Lu,Nd)124] was performed by the KOH flux method. Starting materials of $Lu_2O_3$, $Nd_2O_3$, $Ba(OH)_2$ (anhydrous) and CuO were mixed in a nominal molar ratio of Lu:Nd:Ba:Cu = $1-x$:$x$:2:4 ($x$ = 0, 0.1, 0.2, 0.3). A mixture with KOH of 70 wt% to the total amount was charged into a lid-covered $Al_2O_3$ crucible. The sample was heated at 700-730ºC for 15-40 hours followed by furnace cooling. The obtained sample was washed with distilled water to remove the residual flux and then dried at 110ºC for 30 minutes.

Powder X-ray diffraction (XRD) measurement was performed for the well pulverized samples using the MiniFlex 600 (Rigaku) with Cu K$\alpha$ radiation. Single crystal XRD measurement was performed using the XtaLab mini (Rigaku) with Mo K$\alpha$ radiation. The crystal structure was solved by the program SHELXT on the WinGX software and refined using the program ShelXle.[13-15] The microstructure and the chemical composition of the single crystals were identified by use of a scanning electron microscope (SEM) equipped with an energy dispersive X-ray analyzer (EDX) using JSM-6010LA (JEOL). Magnetic susceptibility was measured for one single crystal covered with a Kapton tape under an applied filed of 50 Oe perpendicular to the $ab$-plane with a superconducting quantum interference device magnetometer (MPMS-XL; Quantum Design). Resistivity was measured for a cluster of the single crystals by the standard four-probe method. Gold wires as four probes were attached on the cluster with silver paste (4922N; Du Pont Co., Ltd.).

Figure 1 shows the powder XRD patterns of (Lu,Nd)124 synthesized at $x = 0$, 0.1, 0.2 and 0.3. The XRD pattern of (Lu,Nd)124 at $x = 0.2$ mainly exhibited (Lu,Nd)124 phase including a few impurities, which were not completely separated from the crystals due to the small single crystals. The diffraction peaks of (Lu,Nd)124 can be indexed on the basis of orthorhombic *Ammm* structure with lattice constants of $a = 3.835(3)$ Å, $b = 3.879(2)$ Å and $c = 27.195(6)$ Å. At $x = 0$, the non-superconducting Lu-related oxides formed. At $x = 0.1$, Lu123 and the Lu-related oxides appeared as major phases and (Lu,Nd)124 was a minor phase. The superconducting transition of Lu123 only was observed at ~88 K in the magnetic susceptibility measurement. At $x = 0.3$, Nd123 formed and (Lu,Nd)124 did not appear.

The single crystals of (Lu,Nd)124 at $x = 0.2$ exhibit a flat plane with 40-120 μm in width and 10-20 μm in thickness, as shown in Fig. 2. The cation ratio of the single crystals normalized by the Cu element was estimated to be Lu:Nd:Ba:Cu = 0.89(6):0.14(3):1.98(2):4 from the EDX analysis, which almost coincides with the nominal composition of Lu:Nd:Ba:Cu = 0.8:0.2:2:4. The single crystal structural analysis was performed and converged on the $R_1$ value of 8.18% and the $wR_2$ value of 19.3% for $I > 2\sigma(I)$ with the goodness of fit of 0.958. Furthermore, the final refinement converged as long as the occupancies of Lu and Nd were the values of 0.91(7) and 0.09(7) (Occ.(Lu) + Occ.(Nd) = 1 at the same site), respectively. Taken account of the EDX result, the Lu site was partially substituted by the Nd ion.

Figure 3 shows the temperature dependence of the magnetic susceptibility for one single crystal of (Lu,Nd)124 at $x = 0.2$. The magnetic susceptibility started to drop at ~75 K and exhibited the perfect diamagnetism throughout the temperature down to 10 K in the zero-field-cooling mode, where the demagnetization effect is excluded.

The superconducting transition was also observed in the resistivity measurement for the cluster of single crystals of (Lu,Nd)124 at $x = 0.2$. The resistivity started to drop at 75 K and reached zero at 65 K, as shown in Fig. 4. The onset temperature of the superconducting transition coincides with the $T_c$ determined from the magnetic susceptibility measurement.

Figure 5 denotes the relationship between the $T_c$ and the $c$-axis length in the RE124 system. The $T_c$ monotonically increases from ~70 K with decreasing the $c$-axis. The $T_c$ and the $c$-axis of (Lu,Nd)124 are on the extension of the $T_c$s in Ref. 4.[4] Compared to the $T_c$s in Ref. 8,[8] those obtained by this work and in Ref. 4 are relatively low but the tendency of the relationship is the same. Therefore, (Lu,Nd)124 successfully represents the $T_c$ and the $c$-axis of Tm124. This means that we can obtain [RE(1),RE(2)]124 equivalent to the one composed of single RE element by combination of two RE elements.

In conclusion, we have synthesized the single crystals of Nd-substituted Lu124 by the KOH flux method. The EDX measurement and the single crystal structural analysis reveal that the compositional ratio of the obtained (Lu,Nd)124 is almost consistent with the nominal composition of $Lu_{0.8}Nd_{0.2}Ba_2Cu_4O_8$ and the Nd ion incorporates into the Lu site. The superconducting transition of (Lu,Nd)124 is observed at ~75 K both in the magnetic susceptibility and the resistivity, which corresponds to Tm124 from the relations between $T_c$ and $c$-axis in RE124 system. Present results indicate that RE124 phase corresponding to the one composed of single RE element can be obtained by combination of two RE elements.

**Acknowledgment(s)**

The authors thank M. Abe, M. Oishi, S. Harada and T. Ishiyama for their supports.

*Corresponding author: HARA.Hiroshi@nims.go.jp

**Figure captions**

Fig. 1. (Color online) Powder XRD patterns of $Lu_{1-x}Nd_xBa_2Cu_4O_8$ at $x$ = 0, 0.1, 0.2 and 0.3.

Fig. 2. SEM image of a typical $Lu_{0.8}Nd_{0.2}Ba_2Cu_4O_8$ single crystal.

Fig. 3. (Color online) Temperature dependence of magnetic susceptibility under field cooling (FC) and zero-field cooling (ZFC) with an external field of 50 Oe perpendicular to the *ab*-plane for one single crystal of $Lu_{0.8}Nd_{0.2}Ba_2Cu_4O_8$. Inset shows the enlargement view near the superconducting transition.

Fig. 4. (Color online) Temperature dependence of resistivity for a cluster of the single crystals of $Lu_{0.8}Nd_{0.2}Ba_2Cu_4O_8$. Right bottom inset shows the enlargement view near the superconducting transition and left upper one shows the optical microscope image of the cluster of the single crystals.

Fig. 5. (Color online) Relationship between $T_c$ and *c*-axis length in RE124 system.

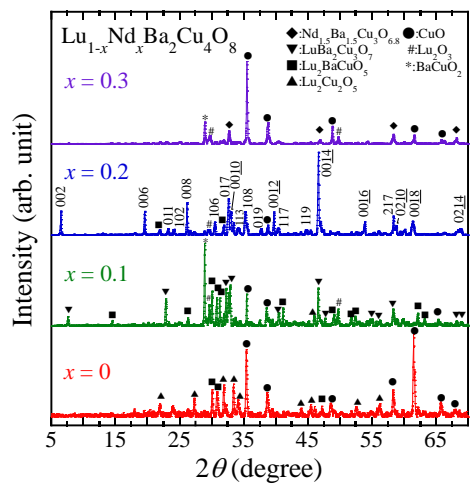

Fig. 1.

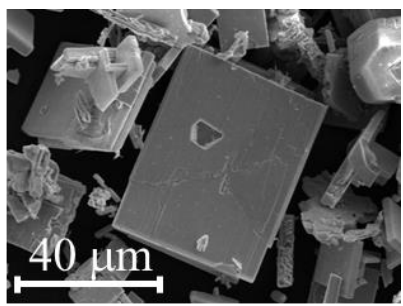

Fig. 2.

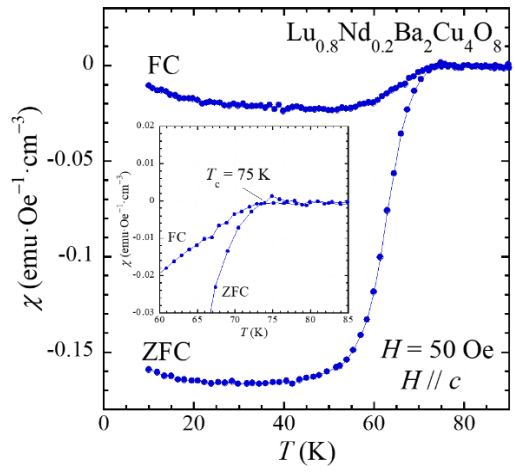

Fig. 3.

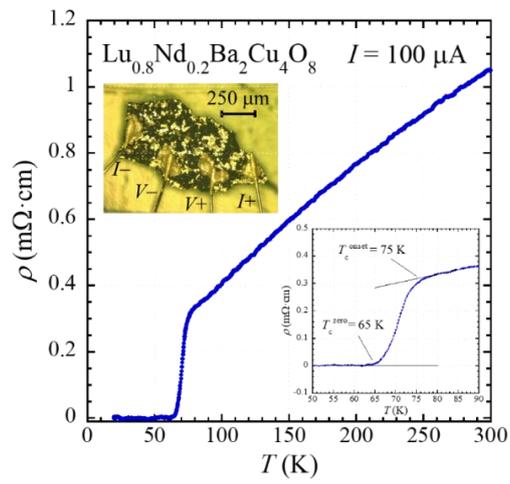

Fig. 4.

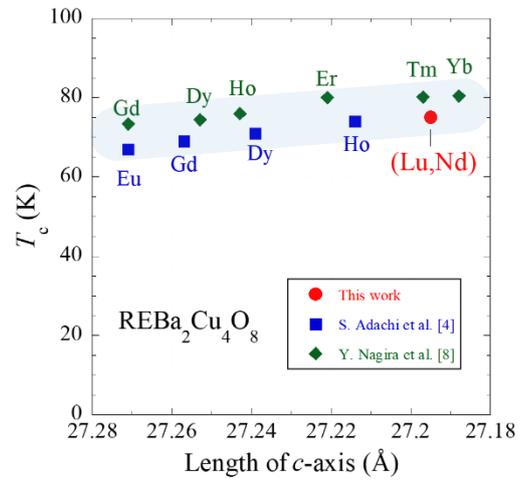

Fig. 5.